\documentclass[final,3p,times,twocolumn]{elsarticle}
\usepackage{graphicx,bm,epsf,float,amssymb}

\usepackage{lineno}
\linenumbers

\journal{Nuclear Instruments and Methods A}

\begin{document}

\begin{frontmatter}
\title{Study of wavelength-shifting chemicals for use in large-scale water Cherenkov detectors}
\author[llnl,davis]{M. Sweany\corref{cor1}}\ead{sweany1@llnl.gov}
\author[llnl]{A. Bernstein}
\author[llnl]{S. Dazeley}
\author[wp]{J. Dunmore \fnref{fn1}}
\author[davis]{J. Felde}
\author[davis]{R. Svoboda}
\author[davis]{M. Tripathi}

\address[llnl]{Lawrence Livermore National Laboratory, Livermore, CA 94550, USA}
\address[davis]{Department of Physics, University of California, Davis, CA 95616, USA}
\address[wp]{Department of Physics and Nuclear Engineering, United States Military Academy, West Point, NY 10996, USA}

\cortext[cor1]{Corresponding Author. Tel: +1 925 424 3732}
\fntext[fn1]{Current Address: Department of Physics, University of Texas at El Paso, El Paso, TX 79968}

\begin{abstract}
Cherenkov detectors employ various methods to maximize light collection at the photomultiplier tubes (PMTs).  These generally involve the use of highly reflective materials lining the interior of the detector, reflective materials around the PMTs, or wavelength-shifting sheets around the PMTs.  Recently, the use of water-soluble wavelength-shifters has been explored to increase the measurable light yield of Cherenkov radiation in water.  These wave-shifting chemicals are capable of absorbing light in the ultravoilet and re-emitting the light in a range detectable by PMTs.  Using a 250 L water Cherenkov detector, we have characterized the increase in light yield from three compounds in water: 4-Methylumbelliferone, Carbostyril-124, and Amino-G Salt.  We report the gain in PMT response at a concentration of 1 ppm as: 1.88 $\pm$ 0.02 for 4-Methylumbelliferone, stable to within 0.5\% over 50 days, 1.37 $\pm$ 0.03 for Carbostyril-124, and 1.20 $\pm$ 0.02 for Amino-G Salt.  The response of 4-Methylumbelliferone was modeled, resulting in a simulated gain within 9\% of the experimental gain at 1 ppm concentration.  Finally, we report an increase in neutron detection performance of a large-scale (3.5 kL) gadolinium-doped water Cherenkov detector at a 4-Methylumbelliferone concentration of 1 ppm.
\end{abstract}


\end{frontmatter}

\section{Introduction} 
Wavelength-shifting (WLS) chemicals have the potential to increase the light response of water Cherenkov detectors by re-emitting the ultraviolet (UV) portion of Cherenkov light into a wavelength for which PMTs have a high quantum efficiency.  A handful of chemicals have been studied previously and were shown to increase the number of detected photons by a factor of around two \cite{dai}.  However, many of these tests have been small table-top studies, not fully deployed detectors; the improvement in such situations is unclear.  Material compatibility, the WLS-doped water attenuation length, UV wall reflectivity, and absorptivity of the chemical all impact the performance in a large-scale detector.  In particular, if the chemical has a negative impact on the water attenuation length, then gain from the UV portion of the Cherenkov spectrum may be lost.  Losses of this nature may not necessarily be observed in small scale tests.

We have performed a series of tests to characterize the gain in light collection as a function of concentration with three chemicals: 4-Methylumbelliferone (4-MU), Carbostyril-124 (CS-124), and Amino-G Salt (AG).  The first series of tests were performed with a 250 L detector, described in Section \ref{sec:wlschar}.  As a final test, 4-MU, the best chemical in terms of light gain, cost, ease of use, and stability was used in a 3.5 kL gadolinium-doped water Cherenkov neutron detector designed for nuclear non-proliferation purposes.  This large-scale detector has been fully characterized without WLS chemicals, and is described in \cite{sweany}.   The neutron detection performance with WLS is described in Section \ref{sec:ndet}.

\section{WLS Characterization}
\label{sec:wlschar}
\begin{figure}\centering
\includegraphics[width=0.5\textwidth]{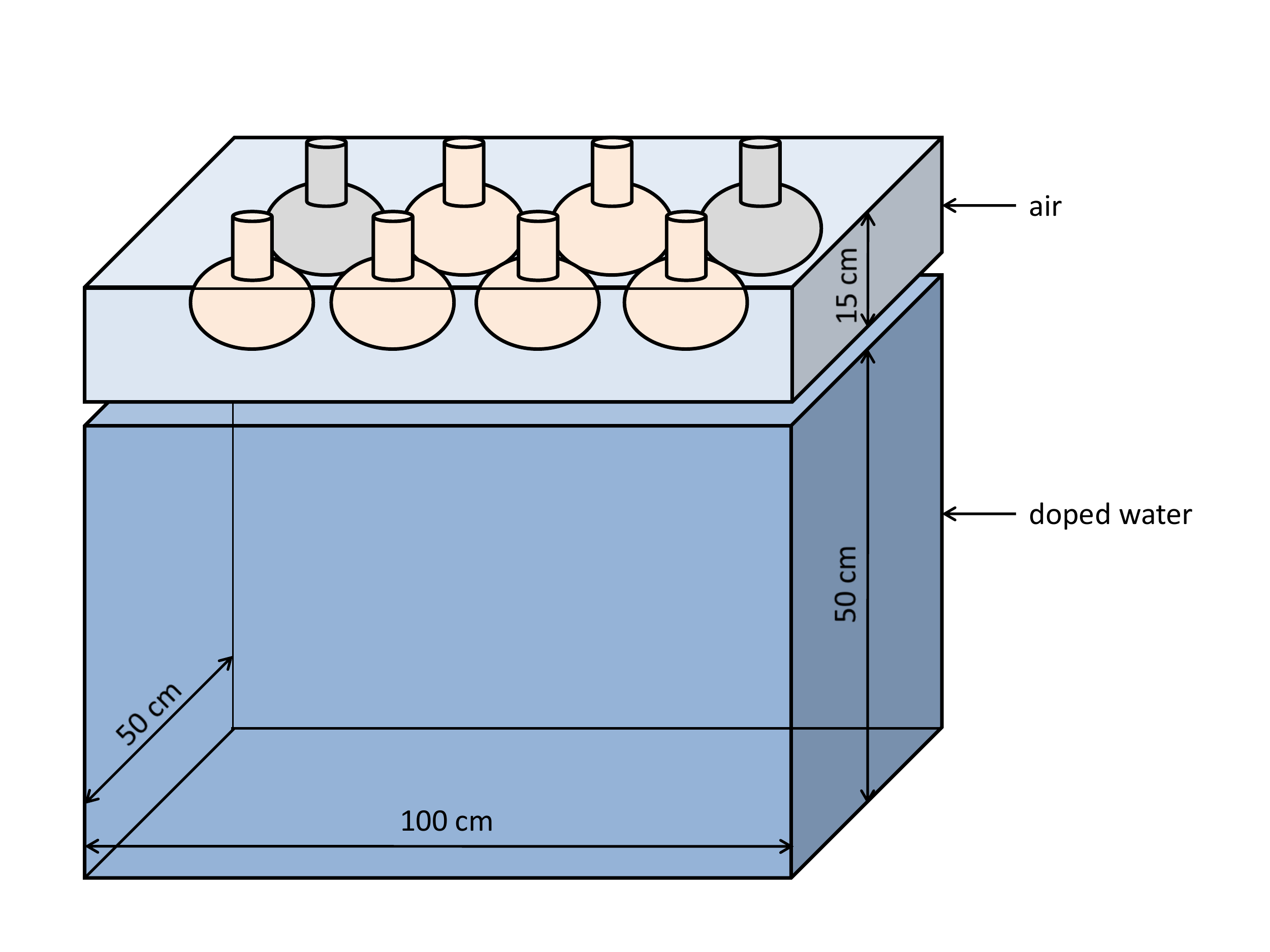}
\caption{Rendering of the 250 liter detector.  Two of the PMTs were nonfunctional, and are grayed-out in this image.}
\label{fig:detector}
\end{figure}

In order to characterize the performance of the chemicals, we have used an existing mid-sized (250 L) water Cherenkov-based detector, described in detail in \cite{dazeley}.  A rendering of this detector is shown in Figure \ref{fig:detector}.  The detector consisted of two separate UV transmitting acrylic tanks: a small tank holding PMTs was placed on top of a larger tank.  An O-ring between the two tanks sealed the volume of the lower tank. The lower tank (1x0.5x0.5 m) contained ultra pure, sterilized water doped with WLS chemical, and constituted the 250 L active target volume of the detector.  It was fitted with a small expansion volume and airlock so that the target remained full, and therefore optically coupled to the top tank despite ambient air pressure variations.  The upper tank contained eight downward facing 8 inch ETL 9354kb PMTs, however two PMTs were nonfunctional at the time of testing. The ETL PMTs have a maximum quantum efficiency ($\sim$30\%) at a wavelength of approximately 350 nm.  Each PMT was shielded from magnetic fields by an 8 inch diameter cylinder of mu-metal.  The walls of the tank were highly reflective in the UV: the tanks were constructed with UV transmitting acrylic, and the outside surface of the acrylic was lined with UV reflective 1073B Tyvek \cite{tyvek1, tyvek2}.  However, the top surface of the detector was not reflective: light that didn't enter a PMT was absorbed.

\begin{figure}\centering
\includegraphics[width=0.5\textwidth]{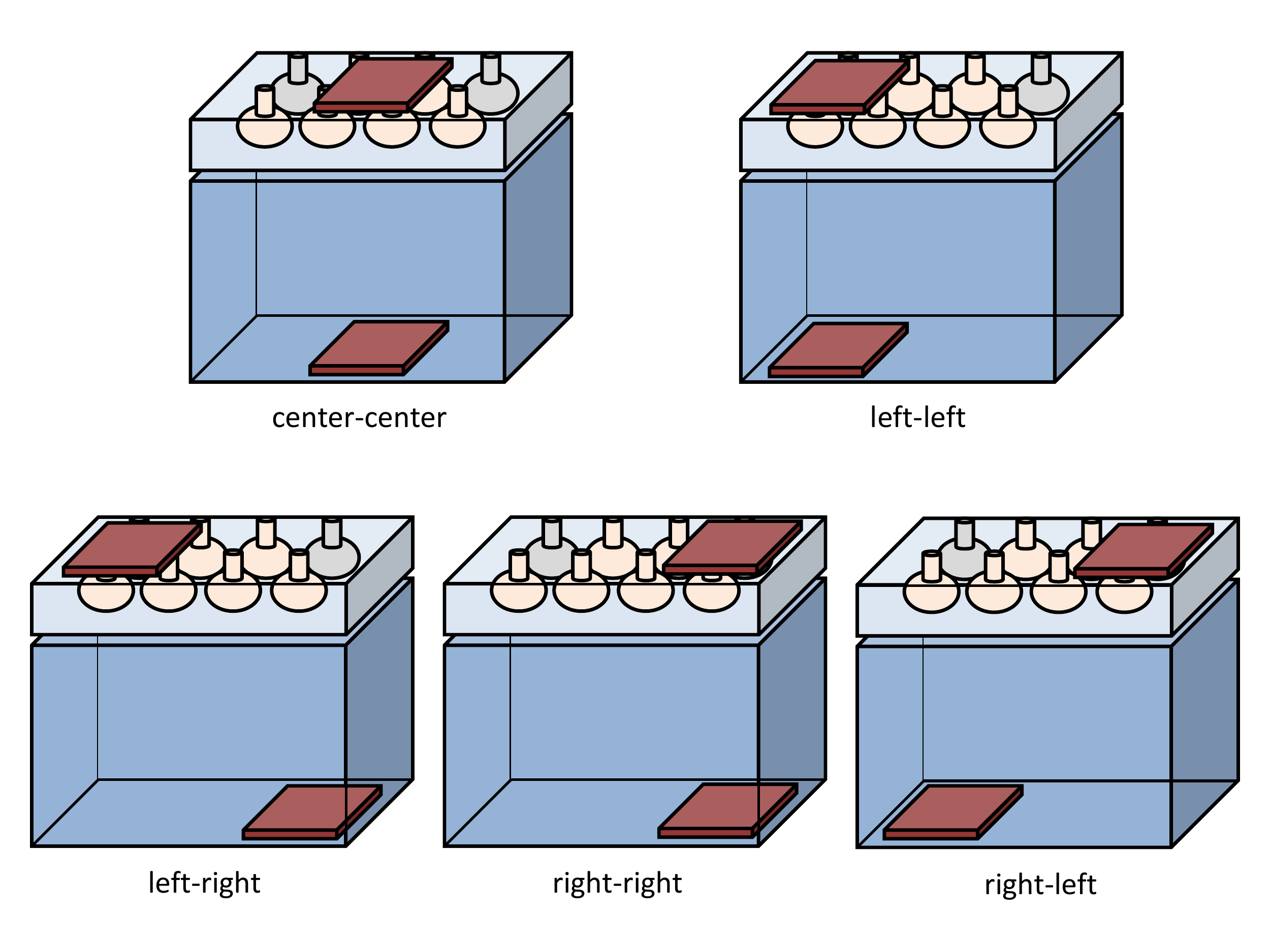}
\caption{The five different paddle configurations, with the top paddle listed first.}
\label{fig:config}
\end{figure}

\subsection{Data Acquisition}
Cherenkov light was measured from cosmic ray muons entering the detector volume.  Two plastic scintillator paddles, each approximately one square foot, were placed above and below the detector volume in various configurations.  The trigger was constructed from a 2-fold coincidence of the paddles.  The detector PMT signals were amplified and shaped at the PMT base with custom built electronics, then underwent additional amplification with a CAEN N568B spectroscopy amplifier before acquisition with a CAEN V785 12-bit ADC.  

\begin{figure}\centering
\includegraphics[width=0.5\textwidth]{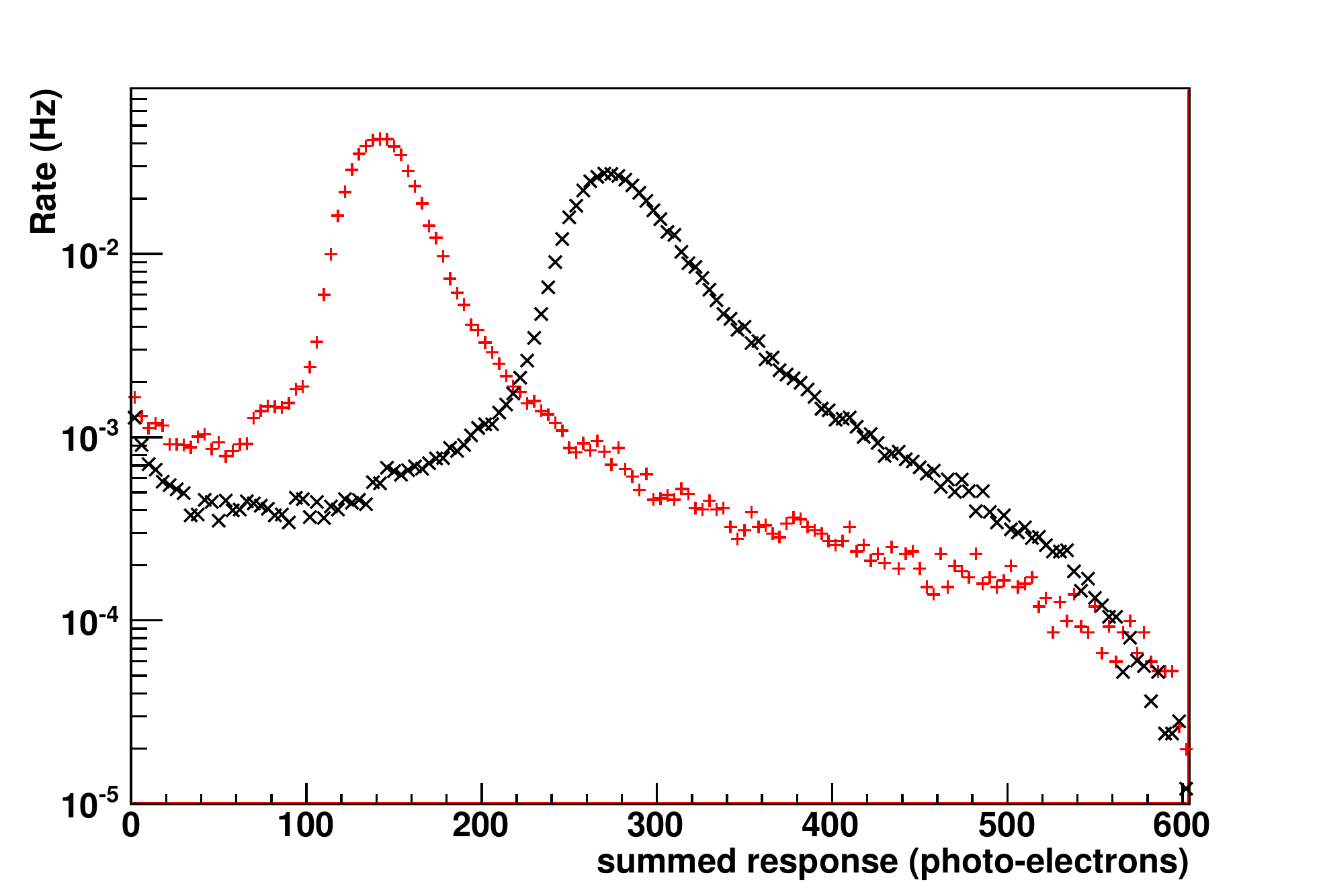}
\caption{The summed PMT response spectrum from muons traversing the detector for pure water alone (red +) and pure water doped with 1 ppm of 4-MU (black x).}
\label{fig:ex}
\end{figure}

Early testing indicated that the individual PMT response was sensitive to the placement of the paddles due to the reflective surfaces in the detector, and the position and orientation of the Cherenkov ring.  Therefore, the PMT response for five different paddle configurations was measured for each WLS concentration; the configurations are shown in Figure \ref{fig:config}.  The data rate was approximately 0.5 Hz for the center-center, left-left, and right-right paddle configurations and 0.2 Hz for the left-right and right-left paddle configurations.  Figure \ref{fig:ex} shows the summed PMT response spectrum from several hours of data in the center-center paddle configuration for both pure water alone, and pure water doped with 1 ppm of 4-MU.

\subsection{Gain and Stability Results}
The stability of both CS-124 and 4-MU (7-hydroxy-4-methylcoumarin) was reported in \cite{dai}.  4-MU was reported to have a pH-sensitive response; however, no statement was made regarding a decrease in emission or absorption for 4-MU at stable pH.  Two other coumarins (4-hydroxycoumarin-3-carboxylic acid and 7-hydroxy-4-methylcoumarin-3-acetic acid) were stated to be unstable within two months at a given pH.  However, stability over two months may be sufficient for certain applications.  In addition, the low cost of 4-MU may warrant scenarios in which the chemical is filtered out and re-applied.  It is not obvious whether the reported variations in the emission and absorption spectrum would be observed, given that PMTs are only sensitive to variations in the integrated absorption spectrum and that a typical PMT quantum efficiency does not vary significantly over the range in the emission spectrum.

\begin{figure}\centering
\includegraphics[width=0.5\textwidth]{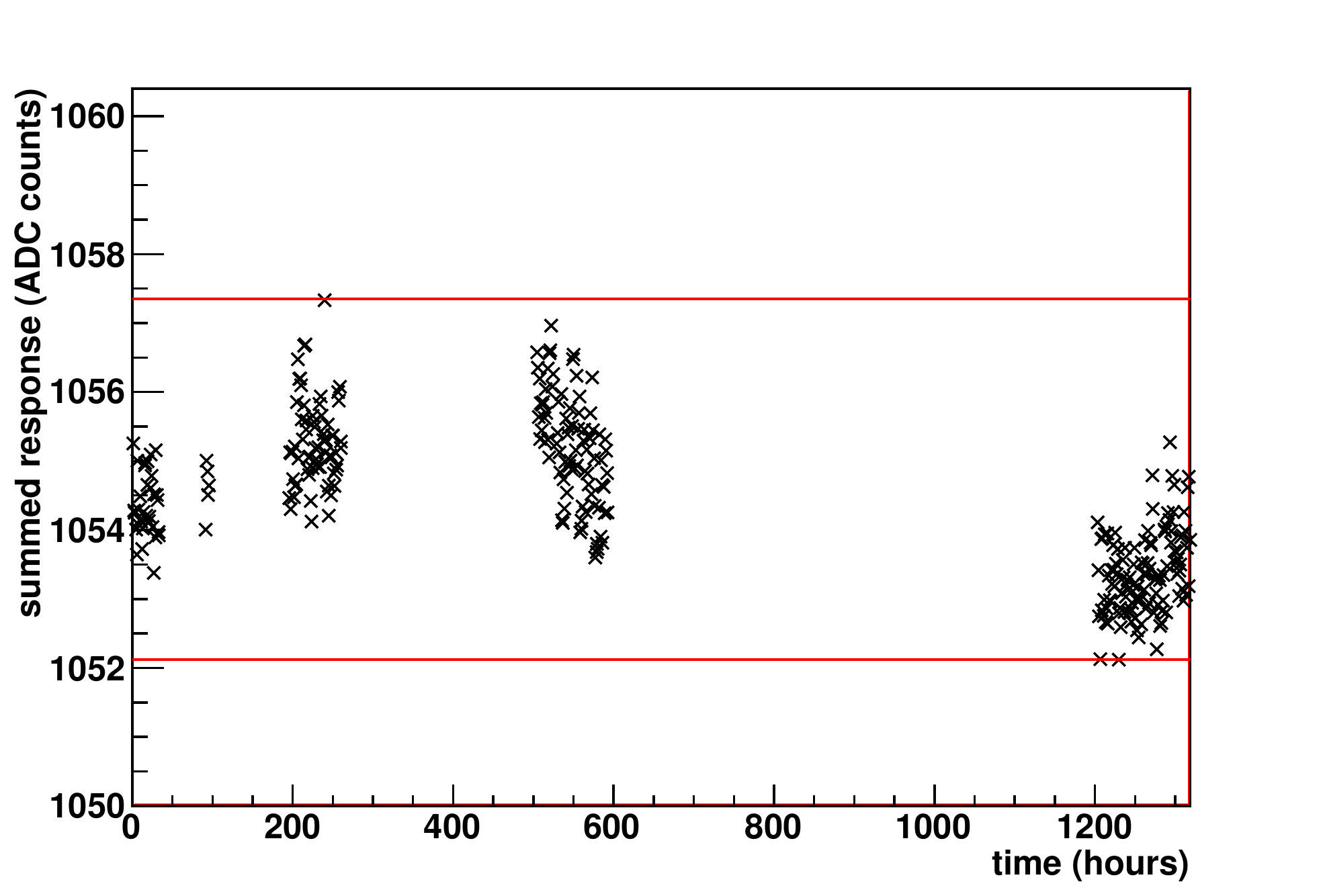}
\caption{The mean of the summed PMT response, measured in ADC counts, for 1 ppm of 4-MU over 50 days of running in the center-center paddle configuration.  The data indicates a systematic spread, shown by the red horizontal lines, of 0.5\%.  Data were taken continuously for several days, with one to two week long delays between runs.}
\label{fig:stability}
\end{figure}

To test stability in PMT response, data were taken in our detector several times over a 50 day period using a concentration of 1 ppm in the center-center paddle configuration. Figure \ref{fig:stability} shows the mean of the summed PMT response for this time period.  The spread of the mean indicates that the PMT response was stable to within 0.5\%.

Five different datasets were acquired for 4-MU in addition to the stability measurement, each consisting of the five muon paddle configurations.  Concentrations of 1/9, 1/3, 1, 3, and 9 ppm were measured.  Calibration measurements with pure water were performed before each dataset and for each paddle configuration to confirm that the overall response of the PMTs was consistent over time.  The gain is defined as the mean of the summed PMT response at concentration divided by the mean of the summed response for the previous water calibration run:

\begin{equation}
gain = \frac{\mu_{wls}}{\mu_{water}}.
\end{equation}

\begin{figure}\centering
\includegraphics[width=0.5\textwidth]{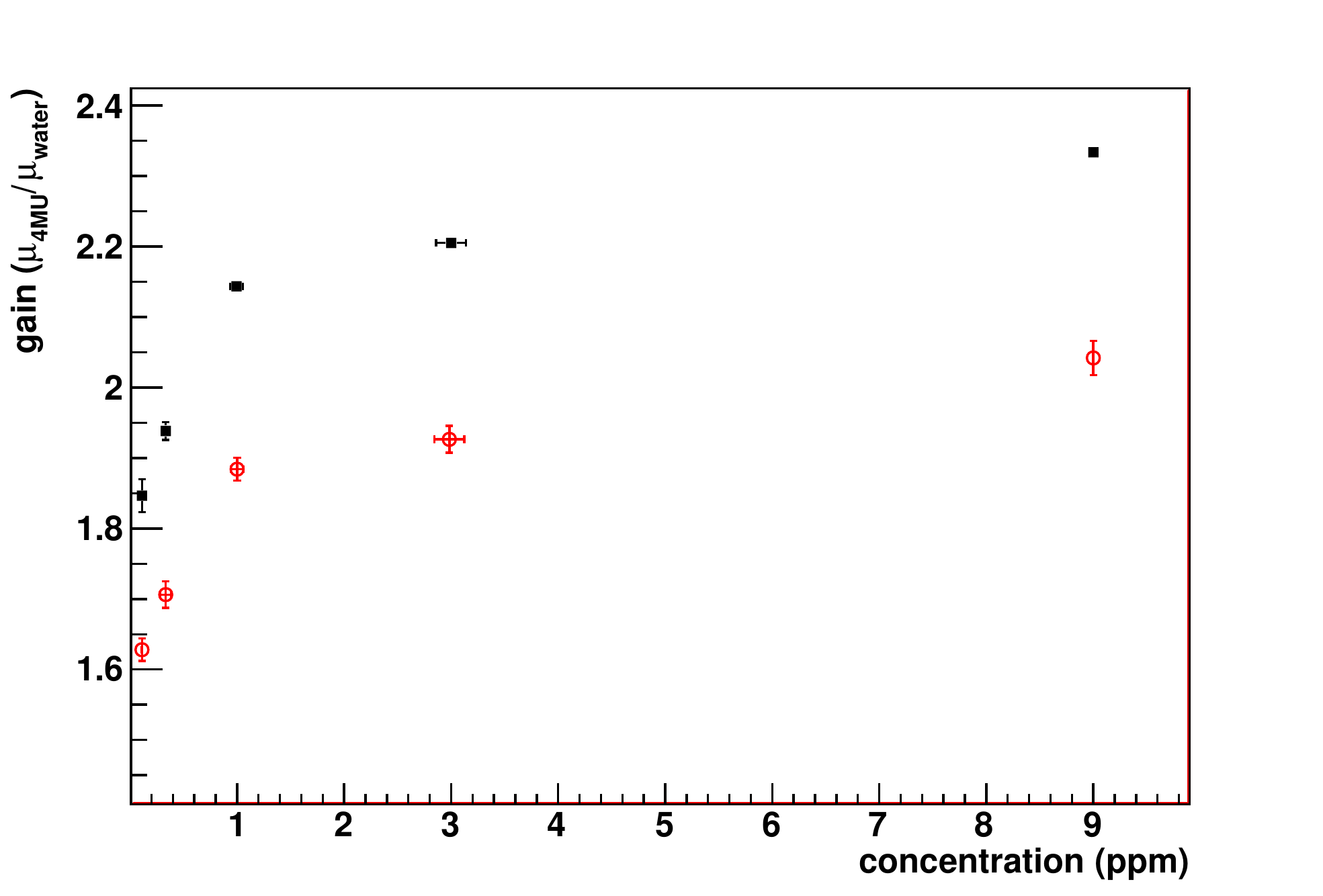}
\caption{The measured gain as a function of concentration for 4-MU.  The two curves are the average of the center-center, left-left, and right-right paddle configurations (red, open circles) and the left-right, right-left paddle configurations (black, filled-in squares).  }
\label{fig:4muGain}
\end{figure}

The two populations of data shown in Figure \ref{fig:4muGain} are due to different Cherenkov light collection efficiencies for muons traveling directly down (center-center, left-left, and right-right) relative to diagonally across the detector (left-right and right-left).  Even though the diagonal muon track lengths were longer, the light collection efficiency was lower.  The addition of WLS caused the light to be emitted isotropically, making up for some of the loss in light collection efficiency.  These two effects combined to form the two populations in the gain curve.  Understanding how the light collection effected the gain measurement was one motivation for performing a detector simulation, described in Section \ref{sec:sim}: these two data populations were observed in our simulation as well.  Although the isotropic nature of the WLS light effected the gain measurement to some degree, the 3.5 kL detector measurement, described in Section \ref{sec:ndet}, provided a more robust measurement in which this effect was washed out by both a top and bottom PMT array.  

The average gain of the center-center, left-left, and right-right configurations was 1.88 $\pm$ 0.02 at 1 ppm.  Because the greatest uncertainty in the measurement was due to variations in the paddle positions causing slightly different average muon path lengths in the detector, we use the spread in the gain within each of the two populations as a measure of the uncertainty.  The gain curve behaved linearly until 1 ppm, then increased very slowly beyond this value.  Saturation is expected to occur because the absorption length, inversely related to the concentration, becomes small compared to the detector size.  
 
\begin{figure}\centering
\includegraphics[width=0.5\textwidth]{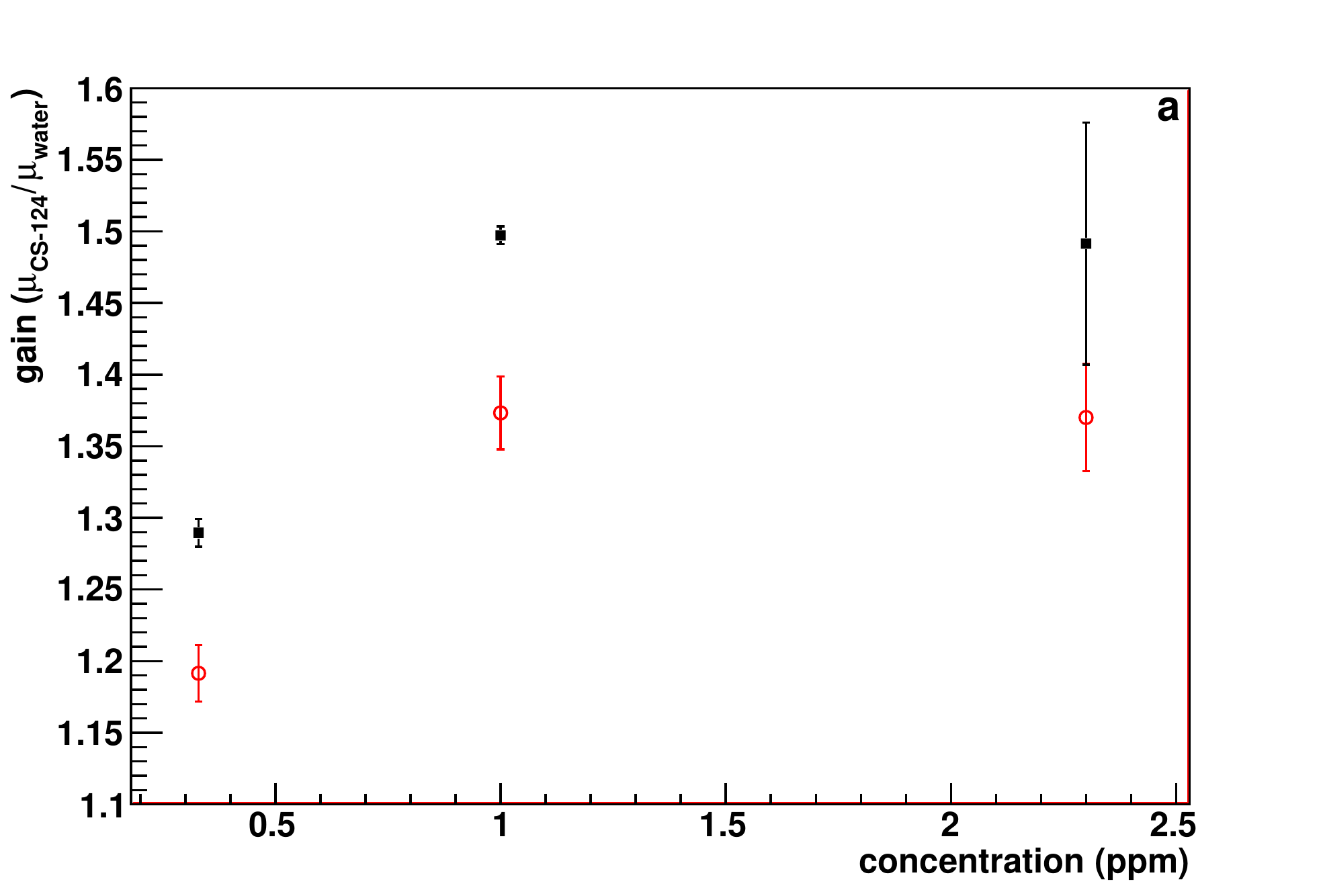}
\includegraphics[width=0.5\textwidth]{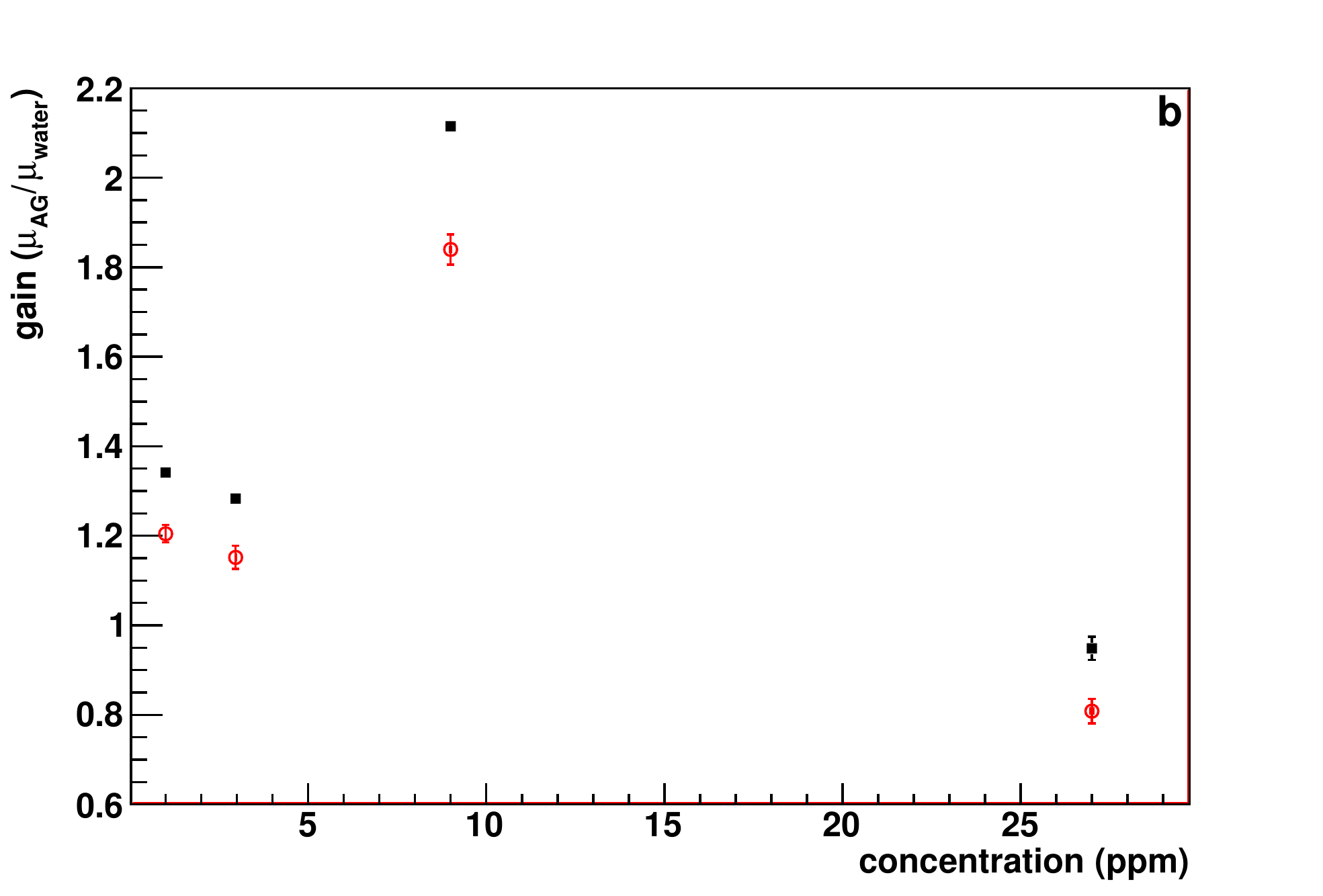}
\caption{The measured gain as a function of concentration for CS-124 (a) and AG (b).  The two curves are the average of the center-center, left-left, and right-right paddle configurations (red, open circles) and the left-right, right-left paddle configurations (black, filled-in squares).  At high concentration (27 ppm), AG caused the water to become brown in color, possibly the cause of a gain value less than one at that concentration. }
\label{fig:CS_AGGain}
\end{figure}

Three datasets were acquired for CS-124 at concentrations of 1/3, 1, and 7/3 ppm, using the same five paddle configurations as for 4-MU.    Figure \ref{fig:CS_AGGain}a shows the average gain for the short and long muon path lengths.  Although the saturation point is not expected to occur at the same concentration as for 4-MU, gains are quoted at 1 ppm as a comparison of how much gain was achieved for the same amount of chemical.  At 1 ppm, the average of the small path length configurations yielded a gain of 1.37 $\pm$ 0.03.

Finally, four concentrations of AG were tested: 1, 3, 9, and 27 ppm.  The results are shown in Figure \ref{fig:CS_AGGain}b.  At 1 ppm, a gain of 1.20 $\pm$ 0.02 was measured for the average of the small muon path length configurations.   The gain vs. concentration curve for this chemical was erratic, and droped below the water response at 27 ppm.  This particular chemical caused the water to become visibly brown in color at the highest concentration (27 ppm), suggesting that the AG adversely affected the optical attenuation length of the mixture.  We do not have an explanation for overall behavior of this chemical.

\subsection{Detector Simulation}
\label{sec:sim}
A full detector simulation in Geant4.9.3 \cite{geant4} was used to characterize the WLS response.  The event generation consisted of muons randomly spread through the two paddle positions.  Cherenkov photons in the 200-600 nm range were generated.  The wall reflectivity and PMT quantum efficiency were matched to the experimental PMT response with no WLS for the center-center configuration, and good overall agreement for the remaining configurations was obtained with the same optical properties.  Then, wavelength shifting properties were added to the water volume.  

Geant4 includes wavelength shifting by adding the following optical properties to a given material: wavelength shifting absorption length ({\scriptsize WLSABSLENGTH}), emission component ({\scriptsize WLSCOMPONENT}), and time constant ({\scriptsize WLSTIMECONSTANT}).  The concentration of the chemical affects the absorption length, and care must be taken not to confuse the Chemist's definition of transmission (log$_{10}$) with that of the Physicist's (log$_e$).  The molar absorptivity, $\epsilon$, with units L mol$^{-1}$ cm$^{-1}$, is reported as a function of wavelength for many WLS chemicals, and related to the transmission, $T$, by 

\begin{equation}
T = 10^{-\epsilon l c},
\end{equation}
\noindent
where $l$ is the path length in a given material and $c$ is the concentration of the absorbing substance.  What Geant4 is expecting for the {\scriptsize WLSABSLENGTH} property is the absorption length $\alpha^{-1}$:

\begin{equation}
T = e^{-\alpha l}.
\end{equation}
\noindent
Equating the two definitions of transmission results in

\begin{equation}
\alpha^{-1} = \frac{1}{\mathrm {ln}(10) \epsilon c} \simeq \frac{1}{2.3 \epsilon c},
\end{equation}
\noindent
relating the absorption length to the given molar absorptivity and concentration.  The concentration $c$ is in units of mol/L.  However, we generally add concentrations in units of mg/L.  Dividing our concentration by the molecular weight of the chemical (in mg/mol) results in the proper concentration units of mol/L.

\begin{figure}\centering
\includegraphics[width=0.5\textwidth]{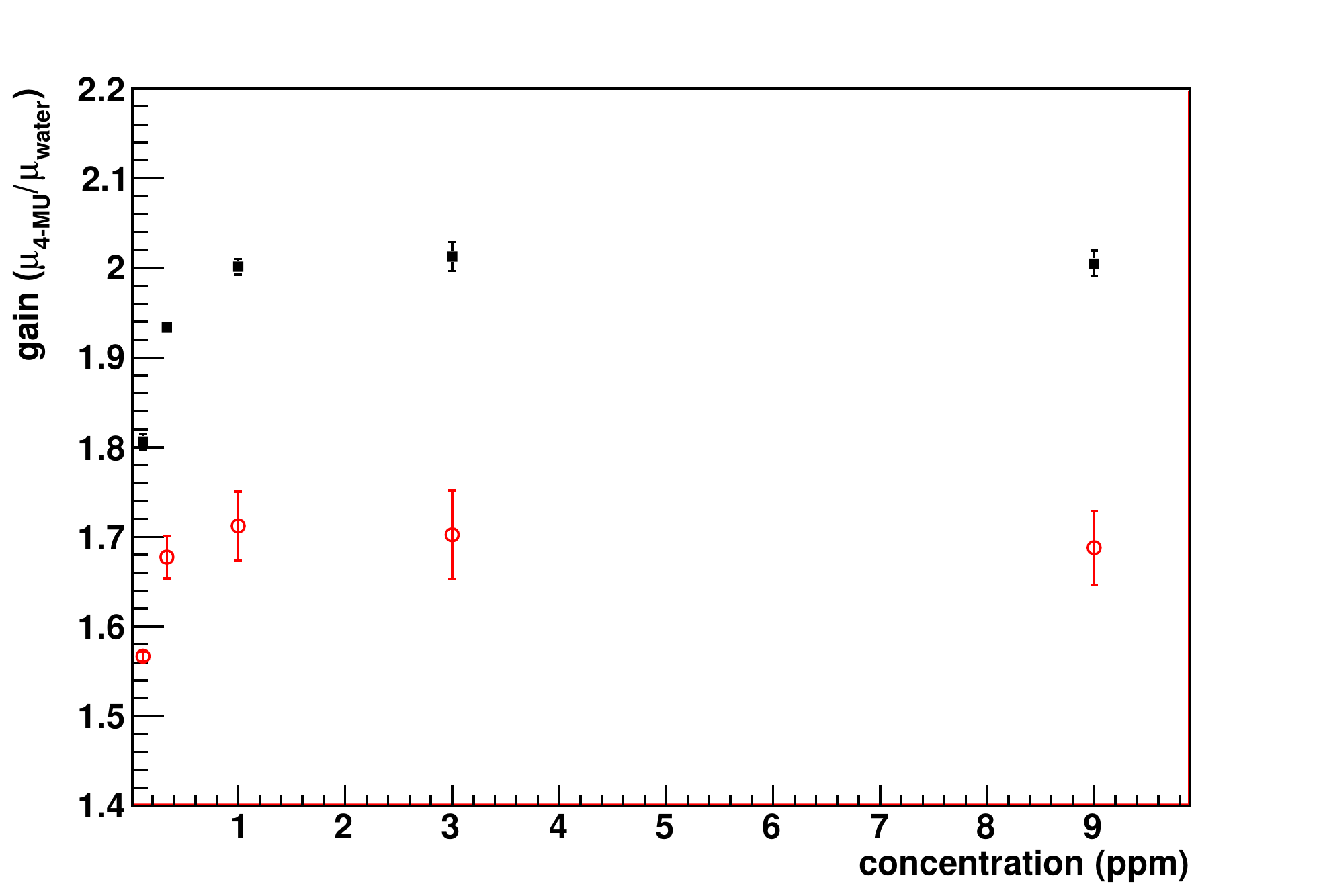}
\caption{The simulated gain as a function of concentration for 4-MU.   The two curves are the average of the center-center, left-left, and right-right paddle configurations (red, open circles) and the left-right, right-left paddle configurations (black, filled-in squares).}
\label{fig:MCGain}
\end{figure}

We know of no applicable wavelength-dependent molar absorptivity and emission spectrum for 4-MU dissolved in water: \cite{snavely} and \cite{abdel} did not report units in their molar absorptivity spectra, and the spectrum below $\sim$ 275 nm was not given.  Since the simulation was only intended to guide usage in the large-scale detector and to understand light collection effects, we approximated the emission spectrum by a Gaussian about the peak value of 450 nm, and the CS-124 molar extinction coefficient was used in place of 4-MU.  After wavelength shifting properties  were assigned to the water, the simulation was used to reproduce the concentration dependent gain.  Figure \ref{fig:MCGain} shows the simulated gain as a function of concentration for 4-MU: the same two populations were reproduced, with saturation occurring near 1 ppm.  The simulation predicted an average gain in the short paddle configurations of 1.71 $\pm$ 0.04 at 1 ppm for 4-MU, within 9\% of our data response.  The error is assumed to result from the imprecise nature of the emission and molar absorptivity spectra.  The simulated gain also has a more obvious saturation at 1 ppm than the data.

\section{Neutron Detection with 4-MU}
\label{sec:ndet}

\begin{figure}\centering
\includegraphics[width=0.5\textwidth]{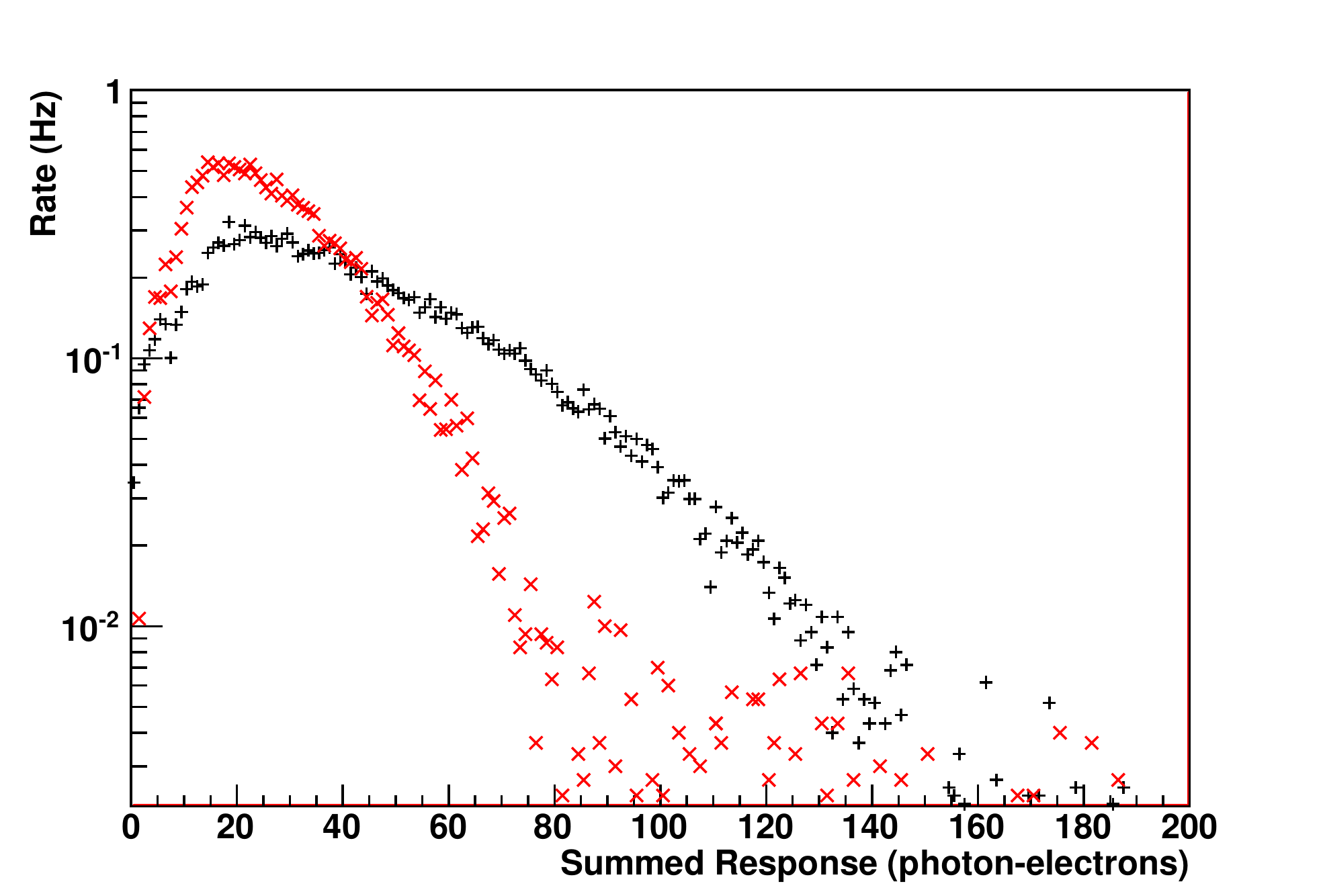}
\caption{The summed PMT response spectrum from $^{252}$Cf neutrons before (red x) and after (black +) adding 1 ppm of 4MU in the 3.5 kL gadolinium-doped water Cherenkov detector.  The position of the high-energy tail increases from around 80 photo-electrons to 160 photo-electrons, indicating a gain in light output of approximately two.}
\label{fig:nSpec}
\end{figure}

After determining 4-MU to be the best candidate in terms of stability, gain, ease of use, and cost, a final measurement was performed with a 3.5 kL gadolinium-doped water Cherenkov detector developed for nuclear non-proliferation purposes, and described in detail in \cite{sweany}.   A comparison of the neutron spectrum from $^{252}$Cf is shown in Figure \ref{fig:nSpec}.  The peak in the spectrum, around 25 photo-electrons, is dominated by the detector threshold: the spectrum is expected to extend down to low energies due to incomplete gamma shower containment.  The position of the high energy tail indicates the energy at which complete gamma shower containment takes place.  As a result of the 4-MU, the high-energy tail of the spectrum extends out to energies approximately twice as high as that without 4-MU, a gain of approximately two as expected.

Previous work \cite{sweany} reported the detected signal over background (S/B) for neutrons emitted from a $^{252}$Cf source positioned at three radial positions outside the detector.  Event level cuts based on timing and energy were established by maximizing the signal significance of neutron rich and neutron poor data sets.  In order to pick out two neutron capture events separated by the characteristic capture time, events were selected based on the charge of both the current and previous event, as well as the time separation of the two.  The S/B was calculated after all cuts from the rates of $^{252}$Cf runs and background runs averaged over 20 seconds.  A 20 second acquisition time was chosen as a determination of the detector's ability to operate under parameters for non-proliferation detection.  Here, we did not restrict ourselves to 20 seconds, and only the rates resulting from the source position 20 cm from the detector wall were measured.  The original measurement without 4-MU from \cite{sweany} is classified as Run 1 in Tables \ref{tab:cuts} and \ref{tab:runs}.

A data run without 4-MU, classified as Run 2, was repeated to determine whether the neutron capture response of the detector changed due to degrading water quality.  Finally, Run 3 was acquired after approximately 1 ppm of 4-MU was added to the detector.  Table \ref{tab:cuts} shows the event level cuts that maximize the signal significance.  Run 1 and 2 show only slight variations in the optimal cut positions.  Run 3 is optimized at a much higher energy, as expected due to the increase in light output from the WLS.

\begin{table}\centering
\begin{tabular}{|r|l|l|l|}
\hline
\bf{Run 1:}				&Parameter			&Left Cut				&Right Cut	\\
\hline
						&Current Charge		& 16 	pe				& 72	pe		\\
						&Previous Charge 		& 16	pe				& 72	pe		\\
						&Inter-event Time		& 4 $\mu$s			& 46	$\mu$s	\\
						&Muon Veto 			&\textgreater46 $\mu$s	& N/A		 \\

\hline\hline
\bf{Run 2:}				&Parameter			&Left Cut				&Right Cut	\\
\hline
						&Current Charge		& 15 	pe				& 69 pe		\\
						&Previous Charge 		& 15	pe				& 69 pe		\\
						&Inter-event Time		& 5 $\mu$s			& 44	$\mu$s	\\
						&Muon Veto 			&\textgreater44 $\mu$s	& N/A		 \\

\hline\hline
\bf{Run 3:}				&Parameter		&Left Cut				&Right Cut	\\
\hline
						&Current Charge	& 26 	pe				& 135 pe		\\
						&Previous Charge 	& 26	pe				& 135 pe		\\
						&Inter-event Time	& 5 $\mu$s			& 40	$\mu$s	\\
						&Muon Veto 		&\textgreater40 $\mu$s	& N/A		 \\
\hline
\end{tabular}
\caption{Analysis cuts obtained by maximizing the signal significance between the background data run and $^{252}$Cf data run with the source seven inches from the detector.  Run 1 is the original run published in \cite{sweany}, Run 2 is a repeat measurement with no 4-MU, and Run 3 was taken after 1 ppm of 4-MU was added to the water. }
\label{tab:cuts}
\end{table}

The S/B is determined from the rates of source and background data averaged over 1k seconds.  The value for Run 1 differs from \cite{sweany} due to the increase in statistics, but are consistent with the errors reported.  The depletion of the source is one expected cause of the decrease between Run 1 and 2.  However, there is a larger decrease than expected from source depletion alone: we attribute this further decrease to long term reductions in water quality.  After the 4-MU was added, the S/B increased from 2.01 $\pm$ 0.03 to 2.40 $\pm$ 0.03.  Table \ref{tab:runs} includes the S/B, as well as the acquisition date of the individual runs.  Between Run 1 and 2, the source depleted to 79\% of its value on October 28th, 2010.  For Run 3, the source depleted to 73\% of the original measurement.  After correcting for source depletion, the S/B increased from 2.54 $\pm$ 0.03 to 3.29 $\pm$ 0.04 between Run 2 and 3.

\begin{table}\centering
\begin{tabular}{r|c|c|c|}
\cline{2-4}
			&Date		& S/B				&Corrected S/B		\\
\cline{2-4}
\bf{Run 1:}	&10/28/10		&2.84 $\pm$ 0.03		&2.84 $\pm$ 0.03	\\
\bf{Run 2:}	&04/29/11		&2.01 $\pm$ 0.02		&2.54 $\pm$ 0.03	\\
\bf{Run 3:} 	&08/08/11		&2.40 $\pm$ 0.03		&3.29 $\pm$ 0.04	\\
\cline{2-4}
\end{tabular}
\caption{The three $^{252}$Cf source runs of interest and their S/B before and after correcting for source depletion.  Run 1 is the original run published in \cite{sweany}, Run 2 is a repeat measurement with no 4-MU, and Run 3 was taken after 1 ppm of 4-MU was added to the water.}
\label{tab:runs}
\end{table}

Finally, several background runs with 4-MU were acquired in order to asses the stability of the PMT response.  No significant difference in the background response was measured over a period of three months.  It would appear that the instability reported in \cite{dai} is not measurable in either of our detectors over a range of two to three months.

\section{Discussion and Conclusions}
We have measured the gain in light yield from muons traversing a 250 L water Cherenkov detector after doping with three different wavelength-shifting chemicals at various concentrations.  Although all three chemicals resulted in a gain, 4-MU resulted in the highest gain, which was stable for nearly two months; it also happens to be the least expensive chemical of the three and dissolves in water without too much difficulty.   CS-124 was particularly difficult to work with: it did not dissolve in the water easily.  In addition, it is the most expensive chemical of the three.  While AG dissolved fairly easily and is moderately priced, it caused the water to become brown at high concentration, and the gain in PMT response with respect to concentration was erratic.  4-MU was chosen as the best chemical to add to our large scale water Cherenkov detector. 

The increase in neutron detection performance of our 3.5 kL gadolinium-doped water Cherenkov due to the addition of 4-MU was measured with a $^{252}$Cf source positioned 20 cm from the detector wall.  We observed an increase in the detector response of approximately two, and the S/B increased from 2.54 $\pm$ 0.03 to 3.29 $\pm$ 0.04.  The gain of two is consistent with our earlier observations of muons traversing the 250 L detector.  After three months in the 3.5 kL detector, no noticeable pitting or degradation occurred on the detector's acrylic or polyethylene components.  Based on these results, we believe 4-MU to be an excellent WLS candidate for water-based Cherenkov detectors: it is stable, dissolves in water, has no noticeable material compatibility issues, and results in a significant improvement in light yield.  In situations in which high water purity is crucial, 4-MU is inexpensive enough that it could be filtered out by a deionizing unit and re-applied.

\section*{Acknowledgments}
The authors wish to thank Dennis Carr for engineering support and Nathaniel Bowden for helpful discussions.  We would also like to thank the DOE NA-22 and the Nuclear Science and Engineering Research Center at the United States Military Academy for their support of this project.  This work was performed under the auspices of the U.S. Department of Energy by Lawrence Livermore National Laboratory under Contract DE-AC52-07NA27344.  Document release number LLNL-JRNL-502992.

\end{document}